# Deformation and failure in nanomaterials via a data driven modelling approach

## M. Amir Siddiq[a,*]

[a]School of Engineering, University of Aberdeen, Fraser Noble Building, AB24 3UE, Aberdeen, United Kingdom

[*]Corresponding Author: amir.siddiq@abdn.ac.uk

## Abstract

A data driven computational model that accounts for more than two material states has been presented in this work. Presented model can account for multiple state variables, such as stresses, strains, strain rates and failure stress, as compared to previously reported models with two states. Model is used to perform deformation and failure simulations of carbon nanotubes and carbon nanotube/epoxy nanocomposites. The model capability of capturing the strain rate dependent deformation and failure has been demonstrated through predictions against uniaxial test data taken from literature. The predicted results show a good agreement between data set taken from literature and simulations.

## 1. Introduction

Material constitutive modelling at various length scales has been under investigation for many decades, however several challenges still exist which are under rigorous research to date. Some of these challenges include, formulation of complex material constitutive models (for e.g. [1]–[6] and references therein) which incorporate underlying physical mechanisms, and identification of a large number of material parameters [7]–[11]. Presently, no single material constitutive model exists which incorporates all physical mechanisms and their interactions. Reasons being complexities associated with the mechanisms and their interactions.

Recently, a number of researchers have started to develop data driven (DD) computing in the context of boundary value problems [12]–[18] and nonparametric regression approach [19]. Such approaches directly use the experimental data and eliminate the efforts, uncertainties and errors induced during inverse modelling to generate stress-strain curves. In the present work, a similar approach is extended and implemented in the context of boundary value problem. As compared to previous works, this research deals with more than two state variables, i.e. stresses, strains, strain rates, and failure, of the nanomaterials.

## 2. Data Driven Computational Framework

The classical formulation has been discussed elsewhere (for details see ref [13]–[15], [18]) and is not repeated here for brevity. A brief summary of the existing model with emphasis on the extension is discussed in the following. Starting with the corresponding phase space for three-dimensional boundary value problem, which comprises of set ($\underline{\sigma}, \underline{\varepsilon}, \underline{\dot{\varepsilon}}, \sigma_f$) of stresses, strains, strain rates, and failure stress, respectively. For the three-dimensional problem the corresponding phase space is assumed to be 19 dimensional with $\underline{\sigma}, \underline{\varepsilon}$, and $\underline{\dot{\varepsilon}}$ are six dimensional each whereas $\sigma_f$ is scalar.

## 2.1 Data Driven Formulation

A discretised finite element model with linear elements of a nonlinear elastic solid is considered as a starting point. Each element ($e$) is comprising of $N$ nodes and $M$ gauss points. The discretised model undergoes displacements $\underline{u}$ which is given by $\underline{u} = \xi_a(x,y,z)\underline{u}_a$ with sum on $a$. Where $\underline{u}_a$ being nodal displacement due to applied nodal forces $\underline{f}_a$, and $\xi_a$ are the interpolation (shape) functions which are based on linear element and are given by

$$\xi_a(x,y,z) = \frac{1}{8}\Sigma^a + \frac{1}{4}x\Lambda_1^a + \frac{1}{4}y\Lambda_2^a + \frac{1}{4}z\Lambda_3^a + \frac{1}{4}yz\Gamma_1^a + \frac{1}{4}xz\Gamma_2^a + \frac{1}{4}xy\Gamma_3^a + \frac{1}{4}xyz\Gamma_4^a \qquad \text{Equation 1}$$

with

$\Sigma^a = [+1 \quad +1 \quad +1 \quad +1 \quad +1 \quad +1 \quad +1 \quad +1]$

$\Lambda_1^a = [-1 \quad +1 \quad +1 \quad -1 \quad -1 \quad +1 \quad +1 \quad -1]$

$\Lambda_2^a = [-1 \quad -1 \quad +1 \quad +1 \quad -1 \quad -1 \quad +1 \quad +1]$

$\Lambda_3^a = [-1 \quad -1 \quad -1 \quad -1 \quad +1 \quad +1 \quad +1 \quad +1]$

$\Gamma_1^a = [+1 \quad +1 \quad -1 \quad -1 \quad -1 \quad -1 \quad +1 \quad +1]$

$\Gamma_2^a = [+1 \quad -1 \quad -1 \quad +1 \quad -1 \quad +1 \quad +1 \quad -1]$

$\Gamma_3^a = [+1 \quad -1 \quad +1 \quad -1 \quad +1 \quad -1 \quad +1 \quad -1]$

$\Gamma_4^a = [-1 \quad +1 \quad -1 \quad +1 \quad +1 \quad -1 \quad +1 \quad -1]$

For a known material dataset, i.e. local phase space ($M_e$), data driven framework searches for optimal local state of each element of the material or structure while at the same time satisfying compatibility and equilibrium, viz.

$$\underline{\varepsilon}_e = \sum_{a=1}^{N} \underline{B}_{ea}\underline{u}_a, \text{ and } \sum_{e=1}^{M} w_e \underline{B}_{ea}^T \underline{\sigma}_e = \underline{f}_a \qquad \text{Equation 2}$$

$\underline{B}_{ea}$ is strain matrix and corresponds to the finite element mesh geometry and connectivity. Based on uniform strain formulation it can be written as

$$\underline{B}_{ea} = \frac{1}{V^e}\int_{V^e} \xi_{a,i}(x,y,z) dV^e \qquad \text{Equation 3}$$

with $\xi_{a,i}(x,y,z) = \frac{\partial \xi_a}{\partial x_i}$ where $i = 1,2,3$.

As mentioned above, material data set ($M_e$) can be comprised of a number of state variables ($\beta_i, i = 1,\ldots,n$). For the present work $n = 19$, i.e. stresses ($\underline{\sigma}$), strains ($\underline{\varepsilon}$), and strain rates ($\underline{\dot{\varepsilon}}$) with six components each, and scalar failure stress ($\sigma_f$), respectively, are known material states.

For the current multi-state data set, following penalty function $F_e$ is used

$$F_e(\beta_i) = \min_{\beta_i' \in M_e} \sum_{i=1}^{n} C_i (\beta_i - \beta_i')^2 \qquad \text{Equation 4}$$

with the minimum is searched for all local states in the data set ($M_e$). Here $C_i$ is a numerical value and does not represent a material property. Overall objective of the solver is to minimise the global $F$ by enforcing conservation law and compatibility constraints as mentioned above (Equation 2)

$$F = \min_{\beta_i' \in M_e} \sum_{e=1}^{M} w_e F_e(\beta_i) \qquad \text{Equation 5}$$

with $w_e$ being the weight factor which is the volume of the element $e$ in undeformed configuration $V_o^e$.

Equations (3) and (4) eliminate the traditional material modelling step which requires material constitutive law, $\underline{\sigma}_e = \hat{\sigma}_e(\underline{\varepsilon}_e, \underline{\dot{\varepsilon}}_e, \sigma_{fe})$, comprising of a number of unknown material parameters which are required to be identified through inverse modelling [8], [9], [11], [20], [21].

Finally, general form of equilibrium constraint is given by

$$\delta\left(\sum_{e=1}^{M} w_e F_e(\sum_{a=1}^{N} \underline{B}_{ea}\underline{u}_a, \underline{\sigma}_e, \underline{\dot{\varepsilon}}_e, \sigma_{fe}) - \sum_{a=1}^{N}\left(\sum_{e=1}^{M} w_e \underline{B}_{ea}^T \underline{\sigma}_e - \underline{f}_a\right)\eta_a\right) = 0 \qquad \text{Equation 6}$$

Following standard procedure of taking possible variations, a system of linear equations is obtained for nodal displacement, the local stresses and the Lagrange multipliers and is given by

$$G^a\left(\underline{u}_b\right) = 0, a, b = 1, \dots, N \qquad \text{Equation 7}$$

Note: For dynamic analyses, inertia can be incorporated separately as

$$M^{ab}\underline{\ddot{u}}_b + G^a\left(\underline{u}_b\right) = 0, a, b = 1, \dots, N \qquad \text{Equation 8}$$

Once all optimal data points are determined, equations 7 are used to define nodal displacements, the local stresses and the Lagrange multipliers.

## 3. Results and Discussions

The applicability of the presented formulation is demonstrated by performing simulations on deformation and failure in carbon-nano-tubes (CNT) and CNT/Epoxy based nanocomposite materials. Brief description of individual experiments, molecular dynamic simulations and results are presented in the following.

### 3.1 Mechanical Response of Carbon Nanotubes

Kok and Wong [23] performed molecular dynamics (MD) studies on single-walled carbon nanotubes (SWCNT) and double-walled carbon nanotubes (DWCNT) to evaluate their mechanical properties. Mechanical properties were estimated for various aspect ratios and strain rates. For the present study, MD results for (5,5) armchair SWCNT at different strain rates were used. For model application purposes and to check the handling of data size by presented model, data for different strain rates was used to generate the stress strain curves for unknown intermediate strain rates (see surface plot in Figure 1 ). At continuum scale, carbon nanotubes have been modelled as shells [24], a beam [25], and a combination of many truss elements [26] (for details and other literature please see references there in). To demonstrate the application of the presented data driven formulation a similar approach has been used, i.e. by modelling the SWCNT as a single truss element under tensile loading. Comparison between MD generated stress strain response and model predictions for three different strain rates are shown in Figure

1. Results show a very good agreement between MD simulations and DD model response for multi-state material dataset.

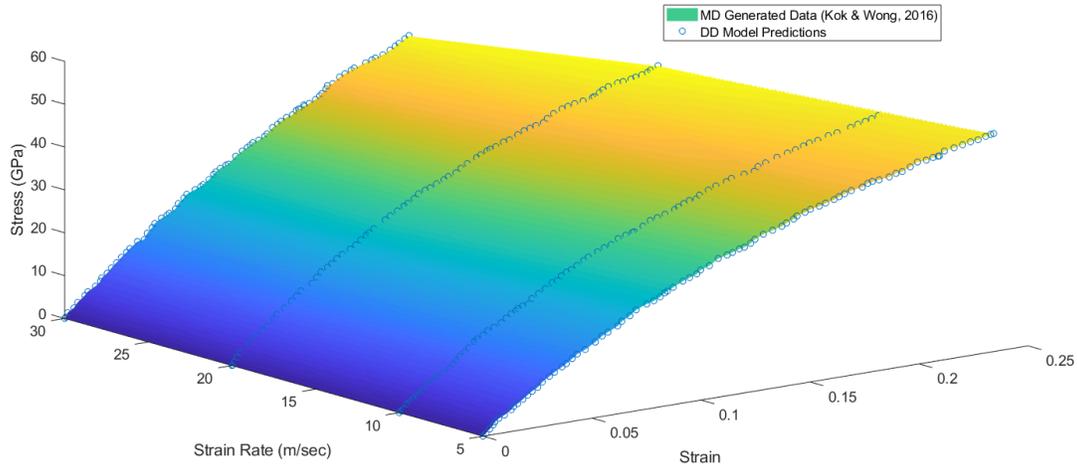

**Figure 1: Comparison of model prediction and MD generated data from Kok and Wong** [23]

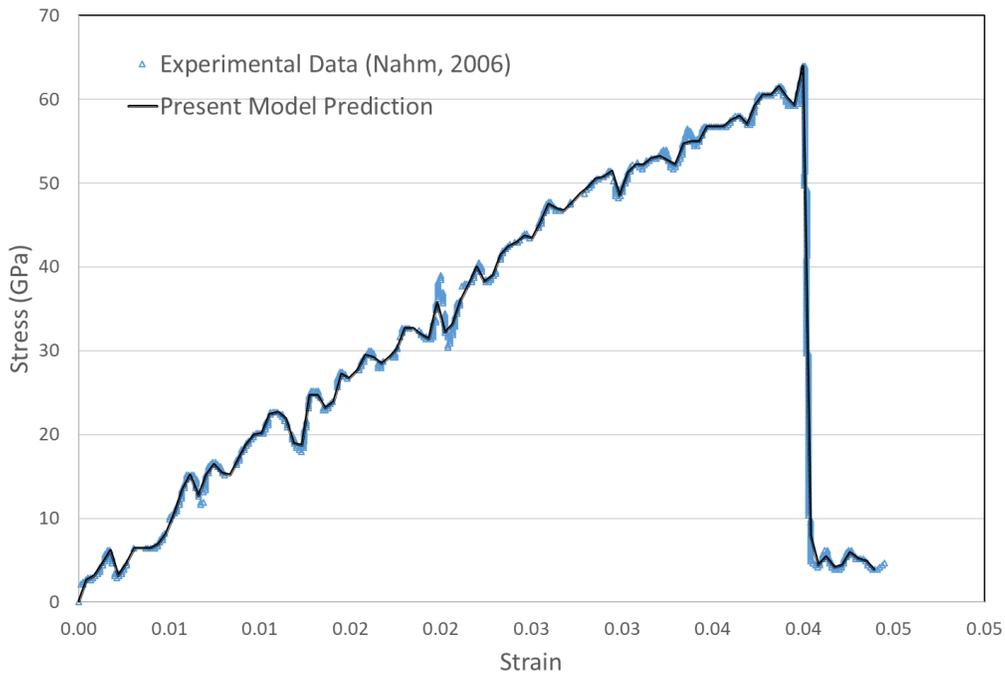

**Figure 2: Comparison of model prediction and experimental data reported in Nahm** [27]

Nahm [27] performed tensile testing of multi-walled carbon nanotubes (MWCNT) using a nano-manipulator and sub nano-resolution force sensor in scanning electron microscope (SEM). As explained above, MWCNT was modelled using a truss element. The results of the comparison of experimental and predicted stress-strain response are plotted in Figure 2 showing a very good agreement up to the final failure of the nanotube.

3.2 Mechanical Response of CNT/Epoxy based Nanocomposites

Yu and Chang [28] performed experimental studies to understand tensile behaviour of MWCNT-reinforced epoxy-matrix composites. Effect of the weight fractions and diameters of CNT's on stress-strain behaviour, and strength was investigated. For the model application purposes, CNT/epoxy composite with 1% MWCNT weight data is used. Uniaxial tension test is simulations using three-dimensional eight node cube element with one integration (gauss) point. A comparison between model predictions and experimental results are presented in Figure 3 showing a good agreement.

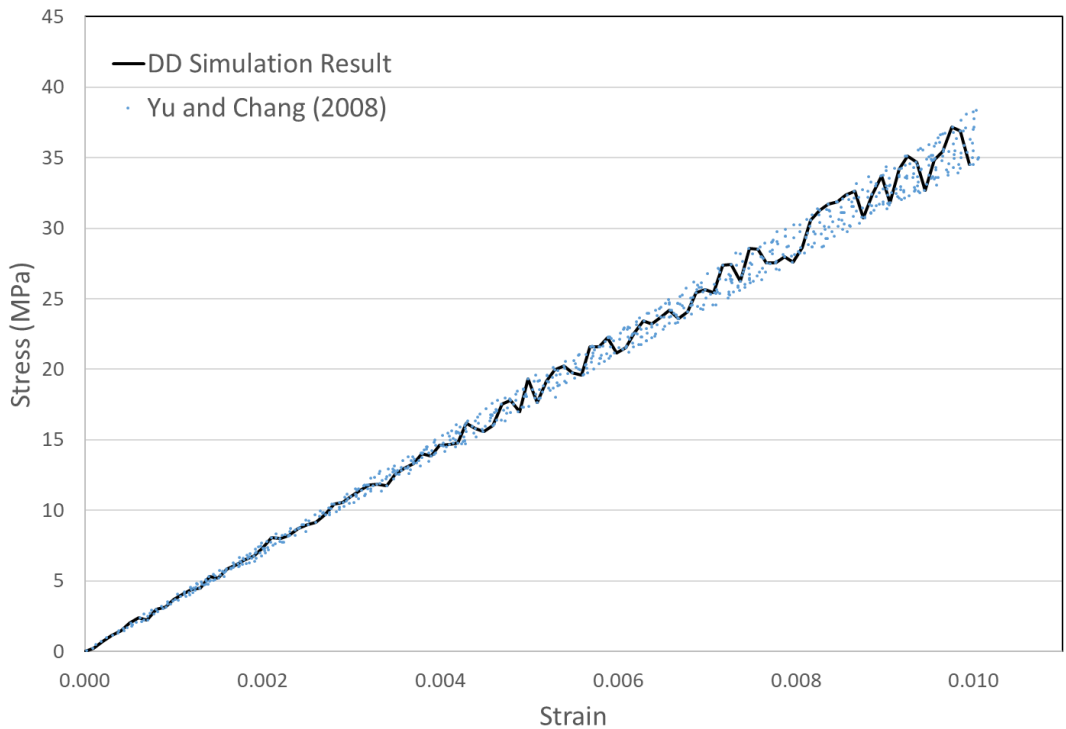

**Figure 3: Comparison of model prediction and experimental data reported in Yu and Chang** [28]

Finally, in order to demonstrate the application of the proposed model to finite element-based analysis and to show the models ability to capture realistic fracture patterns during deformation; a finite element model of dog-bone sample is used. The sample is based on ASTM D638 which was used by Yu and Chang [28] in above presented example. Experimental data from Figure 3 is directly used without using any material constitutive model. Model was discretised using 4488 reduced integration eight node hex elements. Displacement boundary condition of 1.0mm/min was prescribed which was used during experiments. Contour plots of the von Mises stress at different stages of the deformation are plotted in Figure 4 showing the presented DD model's capability of predicting the stresses and macroscopic fracture without any numerical difficulties.

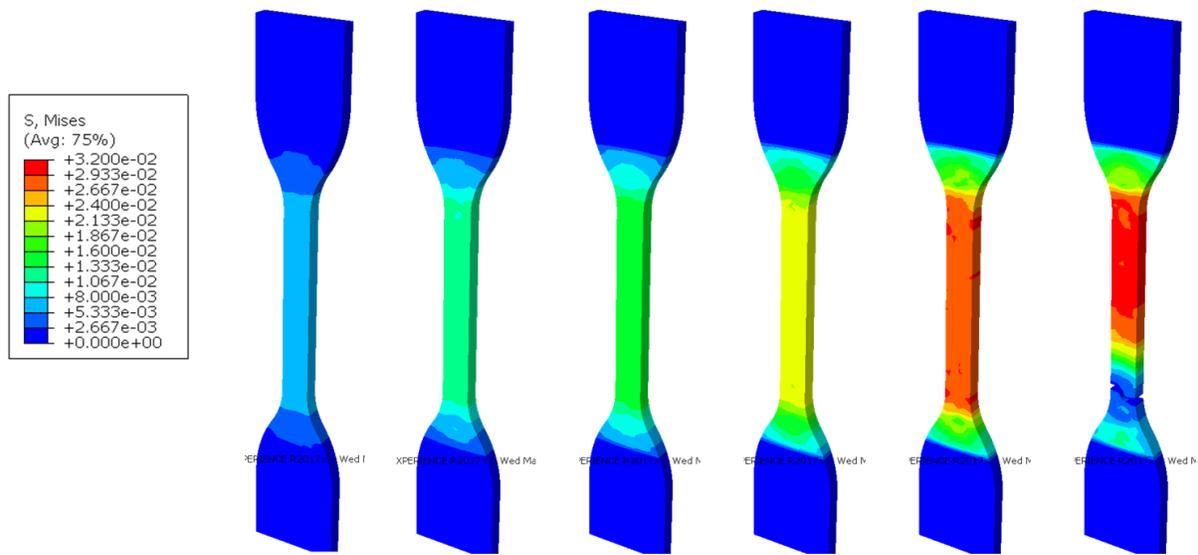

Figure 4: Mechanical response and failure in CNT/Epoxy Composite with 1%wt fraction of CNT under uniaxial tensile loading

## 4. Conclusions

A data driven parameter free model for predicting mechanical response for nanomaterials was presented. Numerical predictions were presented based on data driven computing approach and showed a good agreement with the data set taken from experiments and molecular dynamics simulations. Presented model is applied in the context of nanomaterials, however it can be applied to any length scale. As a future work, the model is further extended to account for plasticity in the context of microscale slip in single crystals. Current crystal plasticity models require a large number of parameters which can be avoided if material data set can directly be used by such models to account for microscale plasticity which is under development and will be reported in near future.

## Acknowledgements

No external funding was received for this project.